\documentclass[preprint,aps,tightenlines,floatfix,showpacs]{revtex4}
\usepackage{graphics}
\usepackage{bm}
\usepackage{amsmath}
\usepackage{amssymb}
\begin{document}

\title
{Simple functional forms for total cross sections from neutron-nucleus 
collisions}
\author{P. K. Deb}
\email{pdeb@mps.ohio-state.edu}
\affiliation
{Department of Physics, The Ohio State University, Columbus, OH 43210, USA.}
\author{K. Amos}
\email{amos@physics.unimelb.edu.au}
\affiliation
{School of Physics, The University of Melbourne, Victoria 3010, Australia}

\date{\today}
\begin{abstract}
Neutron scattering total cross sections have been estimated  
from nuclei ranging in mass 6 to 238 and for projectile energies 10 MeV to 
600 MeV using a simple function of three parameters. These total cross sections
 have also been calculated using coordinate space optical 
potentials formed by full folding effective two-nucleon ($NN$) interactions
with one body density matrix elements (OBDME) of the nuclear ground states.
 Adjusting the theoretical defined parameter values has
enabled us to fit the actual measured data. The simple functional parameter 
values vary smoothly with mass and energy. 
\end{abstract}
\pacs{25.40.-h,24.10.Ht,21.60.Cs}
\maketitle

\section{Introduction}

Neutron-nucleus scattering total cross sections  are required in 
many important  fields of study~\cite{De04a}. Scattering complements
structure theories and probes the matter distributions. It gives relative
motion wave functions which are needed for analyses of other processes;
inelastic, charge exchange, particle transfer, capture reactions etc. In 
Astrophysical studies, reaction cross sections are input to the solar model, 
supernova evolution (formation and reactions of exotic nuclei) and cosmic 
model (nucleosynthesis at t $\sim$ 3 min). Total cross sections are also needed in
disposal of radioactive waste, where using spallation processes, highly 
radioactive elements are transformed to limit radioactivity to acceptable 
time periods. In radiation therapy and radiation protection, dosimetry 
relies on neutron cross sections. Neutron cross sections needed especially for 
(a) H, C, and O (most abundant in body tissue), (b) Si (shielding material and 
detectors), (c) N and P (present in tissue and bones), (d) Ca 
 (present in bones), and 
(e) Al, Fe, Cu, W and Pb (collimation, beam shaping, shielding). In 
medical radiotherapy absorbed dose distributions in the patient are needed and 
cannot be measured directly, they must be calculated. For all these purposes an extensive data
bank is necessary. Since we do not have the experimental data for 
all necessary nuclei at all different energies it would be utilitarian if such 
scattering data were
well approximated by a simple convenient function form with which predictions 
could be made for cases of energies and/or masses as yet to be measured.  
It has been shown~\cite{Ma01,Am02,Ken02,De03} that such forms may exist for 
proton total reaction cross sections.  Herein we consider that concept further 
to reproduce the measured total 
cross sections from neutron scattering for 
energies to 600 MeV and from nine nuclei ranging in mass between ${}^6$Li and 
${}^{238}$U. And we show  that there is a simple three parameter 
function form one can use to form estimates without recourse to 
optical potential calculations. These suffice to show that such forms will also be applicable in 
dealing with other stable nuclei since their neutron total cross sections vary
so similarly with energy~\cite{Ko03}.

%%%%%%%%%%%%%%%%%%%%%%%%%%%%%%%%%%%%%%%%%%%%%%%%%%%%%%%%%%%%%%%%%%%%%%%%%%%%%
\section{ Formalism}

The total and total reaction cross sections for nucleons scattering from nuclei can be expressed 
in terms of partial wave scattering ($S$) matrices specified at energies 
$E\propto k^2$, by
%%%%%%%%%%%%%%%%%%%%%%%%%%%%%%%%%%%%%%%%%%%%%%%%%%%%%%%%%%%%%%%%%%%%%%%%%%%%%
\begin{equation}
S^{\pm}_l \equiv S^{\pm}_l(k) = e^{2i\delta^{\pm}_l(k)} =
\eta^{\pm}_l(k)e^{2i\Re\left[ \delta^{\pm}_l(k) \right] }\ ,
\end{equation}
%%%%%%%%%%%%%%%%%%%%%%%%%%%%%%%%%%%%%%%%%%%%%%%%%%%%%%%%%%%%%%%%%%%%%%%%%%%%%
where $\delta^\pm_l(k)$ are the (complex) scattering phase shifts and 
$\eta^{\pm}_l(k)$ are the moduli of the $S$ matrices. The superscript 
designates $j = l\pm 1/2$. In terms of these quantities, the elastic, reaction
(absorption), and total cross sections respectively are given by
%%%%%%%%%%%%%%%%%%%%%%%%%%%%%%%%%%%%%%%%%%%%%%%%%%%%%%%%%%%%%%%%%%%%%%%%%%%%%
\begin{eqnarray}
\sigma_{\text{el}}(E) & = & \frac{\pi}{k^2} \sum^{\infty}_{l = 0} \left\{
\left(l + 1 \right)\left|S^+_l(k) - 1 \right|^2 + l\left|S^-_l(k) - 1\right|^2 
\right\} = \frac{\pi}{k^2} \sum_l \sigma_l^{(el)}\\
\sigma_{\text{R}}(E) & = & \frac{\pi}{k^2} \sum^{\infty}_{l = 0}\left\{ \left(
l + 1 \right) \left[ 1 - \eta^+_l(k)^2 \right] + l \left[ 1 - \eta^-_l(k)^2 
\right] \right\} = \frac{\pi}{k^2} \sum_l \sigma_l^{(R)}\ ,
\label{xxxx}
\end{eqnarray}
%%%%%%%%%%%%%%%%%%%%%%%%%%%%%%%%%%%%%%%%%%%%%%%%%%%%%%%%%%%%%%%%%%%%%%%%%%%%%
and
%%%%%%%%%%%%%%%%%%%%%%%%%%%%%%%%%%%%%%%%%%%%%%%%%%%%%%%%%%%%%%%%%%%%%%%%%%%%%
\begin{eqnarray}
\sigma_{\text{TOT}}(E) & = & \sigma_{\text{el}}(E) + \sigma_{\text{R}}(E)
= \frac{\pi}{k^2} \left[\sigma_l^{(el)} + \sigma_l^{(R)}\right] 
= \frac{2\pi}{k^2} \sum_l \sigma_l^{(TOT)}\ ,
\nonumber\\ 
\sigma_l^{(TOT)} & = &  \left( l + 1 \right) 
\left\{ 1 - \eta^+_l(k)\cos\left( 2\Re\left[ \delta^+_l(k) \right] \right) 
\right\} + l\left\{1 - \eta^-_l(k) \cos\left( 2\Re\left[ \delta^-_l(k) \right] 
\right) \right\}\ .
\label{SumTOT}
\end{eqnarray}
%%%%%%%%%%%%%%%%%%%%%%%%%%%%%%%%%%%%%%%%%%%%%%%%%%%%%%%%%%%%%%%%%%%%%%%%%%%%%
Therein the $\sigma_l^{(X)}$ are defined as partial cross sections of the total
elastic, total reaction, and total scattering itself.   For proton scattering, 
because Coulomb amplitudes diverge at zero degree scattering, only total 
reaction cross sections are measured. Nonetheless study of such 
data~\cite{Am02,De03} established that partial total reaction cross sections 
$\sigma_l^{(R)}(E)$ may be described by the simple function form
%%%%%%%%%%%%%%%%%%%%%%%%%%%%%%%%%%%%%%%%%%%%%%%%%%%%%%%%%%%%%%%%%%%%%%%%%%%%%
\begin{equation}
\sigma_l^{(R)}(E) = (2l+1) \left[1 + e^{\frac{(l-l_0)}{a}}\right]^{-1} + 
\epsilon\ (2l_0 + 1)\ e^{\frac{(l-l_0)}{a}} \left[1 + e^{\frac{(l-l_0)}{a}}
\right]^{-2}\ ,
\label{Fnform}
\end{equation}
%%%%%%%%%%%%%%%%%%%%%%%%%%%%%%%%%%%%%%%%%%%%%%%%%%%%%%%%%%%%%%%%%%%%%%%%%%%%%
with the tabulated values of $l_0(E,A)$, $a(E,A)$, and $\epsilon(E,A)$ all 
varying smoothly with energy and mass.  Those studies were initiated with the
partial reaction cross sections determined by using complex, non-local, 
energy-dependent, optical potentials generated from a $g$-folding 
formalism~\cite{Am00}. While those $g$-folding calculations did not always give
excellent reproduction of the measured data (from $\sim$ 20 to 300 MeV for 
which one may assume that the method of analysis is credible), they did show a 
pattern for the partial reaction cross sections that suggest the simple 
function form given in Eq.~(\ref{Fnform}).  With that form excellent 
reproduction of the proton total reaction cross sections for many targets and
over a wide range of energies were found with parameter values that varied 
smoothly with energy and mass.

Herein we establish that total cross sections for scattering of 
neutrons from nuclei can also be so expressed and we suggest forms, at least
first average result forms, for the characteristic energy and mass variations 
of the three parameters involved. Nine nuclei, $^6$Li, $^{12}$C, $^{19}$F, 
$^{40}$Ca, $^{89}$Y, $^{184}$W, $^{197}$Au, $^{208}$Pb and $^{238}$U, for
which a large set of experimental data exist, are considered.

%%%%%%%%%%%%%%%%%%%%%%%%%%%%%%%%%%%%%%%%%%%%%%%%%%%%%%%%%%%%%%%%%%%%%%%%%%%%%
\section{ Results and discussions}

\begin{figure}
\centering
\scalebox{0.7}{\includegraphics{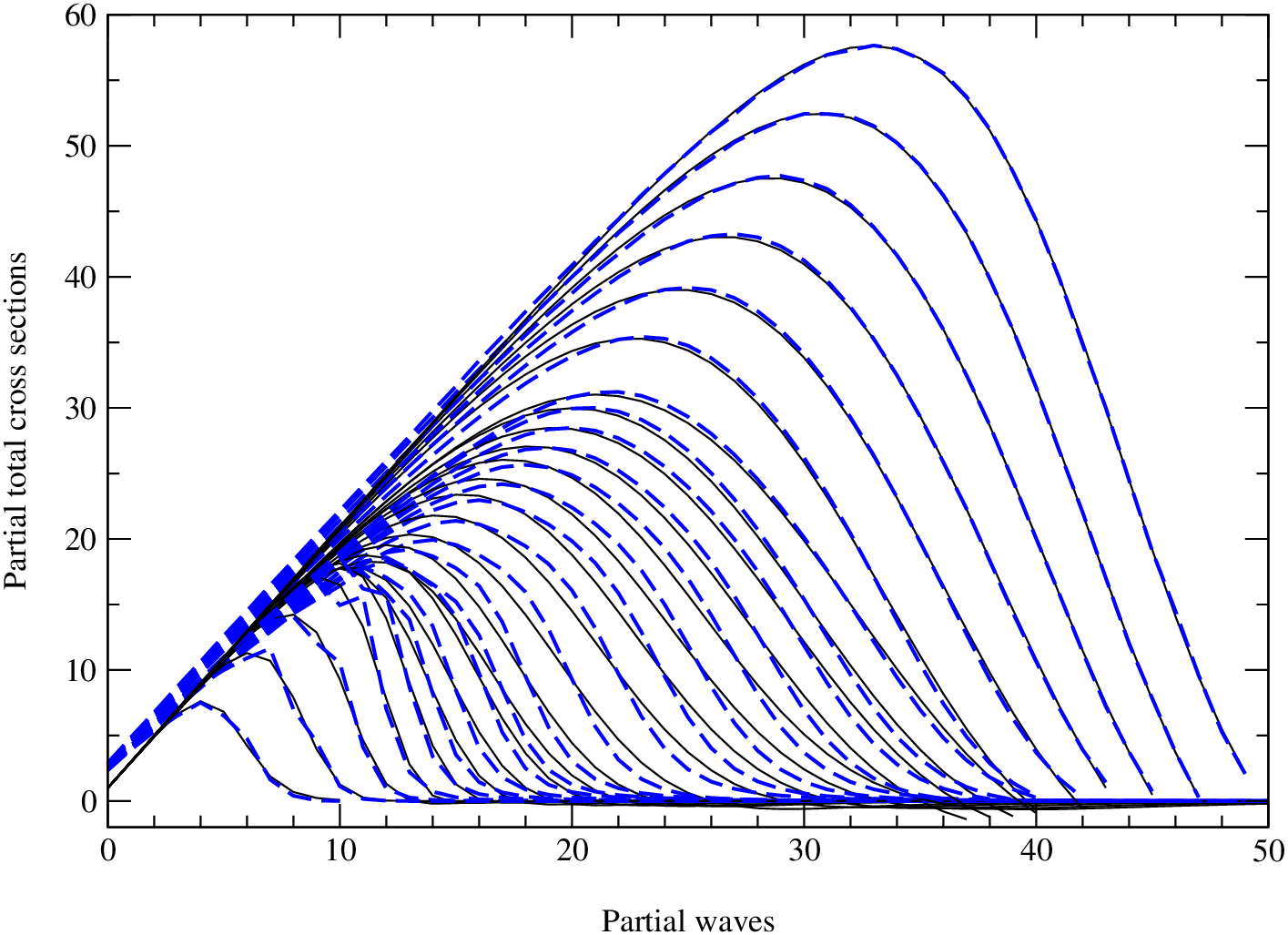}}
\caption{\label{Pb208-partials}
The partial total cross sections for scattering of neutrons from ${}^{208}$Pb
with the set of energies between 10 and 600 MeV specified in the text. The
largest energy has the broadest spread of values.}
\end{figure}

While we have used the partial total cross sections from microscopic results for
neutron scattering from all the nine nuclei chosen and at all of the energies
indicated, only those obtained for ${}^{208}$Pb are shown in
Fig.~\ref{Pb208-partials}. The results from calculations of scattering from
the other eight nuclei have similar form.  The `data' shown as blue dashed
lines in Fig.~\ref{Pb208-partials} are the values found
from the $g$-folding optical model calculations. Each black curve shown therein
is
the result of a search for the best fit values of the three parameters, $l_0$,
$a$, and $\epsilon$ that map Eq.~(\ref{Fnform}) (now for total neutron cross
sections) to these `data'.
%%%%%%%%%%%%%%%%%%%%%%%%%%%%%%%%%%%%%%%%%%%%%%%%%%%%%%%%%%%%%%%%%%%%%%%%%%%%%
%%%%%%%%%%%%%%%%%%%%%%%%%%%%%%%%%%%%%%%%%%%%%%%%%%%%%%%%%%%%%%%%%%%%%%%%%%%%
 
\begin{figure}
\centering
\scalebox{0.7}{\includegraphics{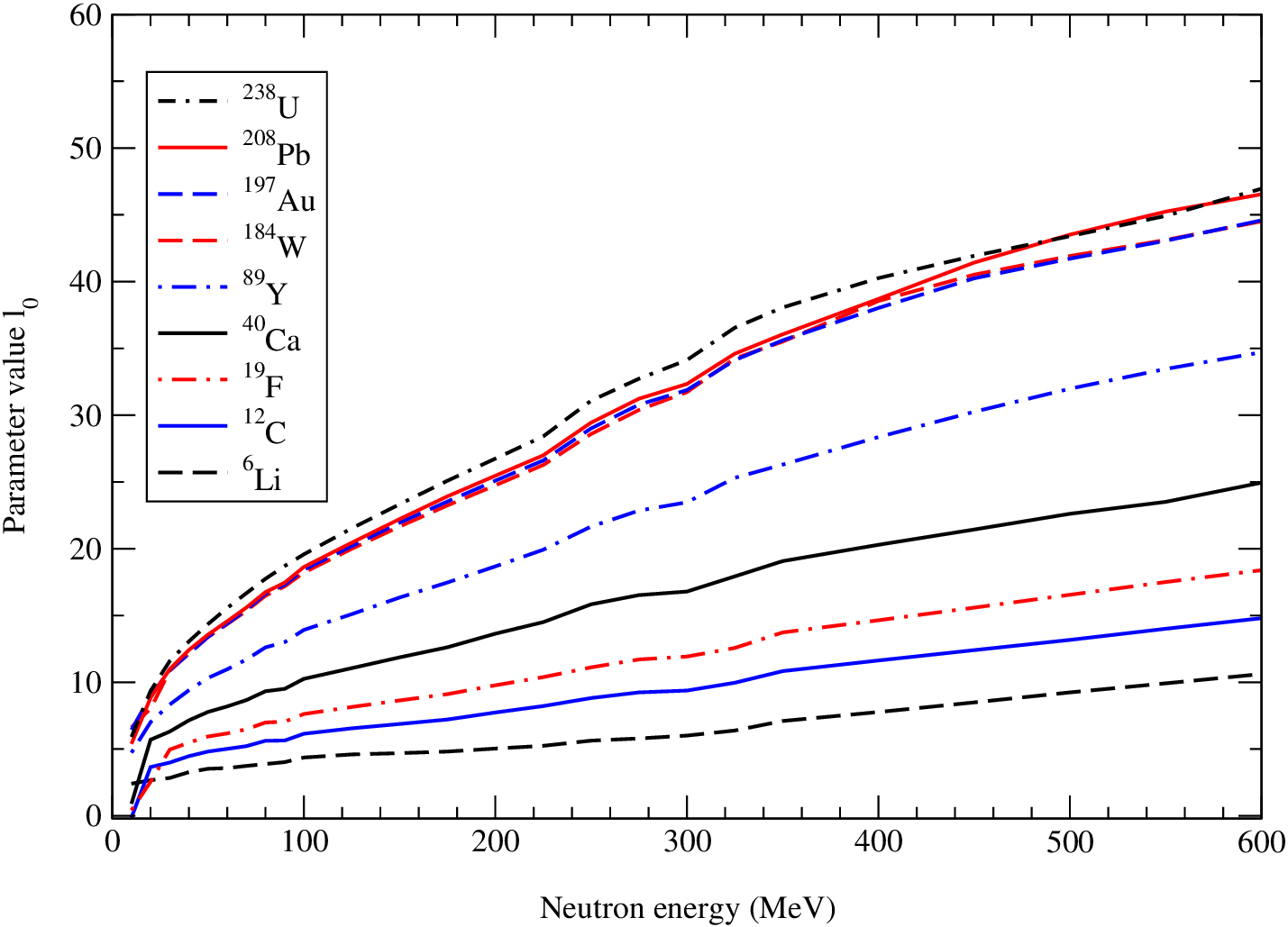}}
\caption{\label{l0vsE} 
The values of $l_0$ that fit neutron total scattering cross-section data from 
the nine nuclei considered and for energies between 10 and 600 MeV. The curves
portray the best fits found by taking a function form for $l_0(E)$.}
\end{figure}

From the sets of values that result from the fitting process, the two 
parameters $a$ and $\epsilon$ can themselves be expressed by the parabolic 
functions 
%%%%%%%%%%%%%%%%%%%%%%%%%%%%%%%%%%%%%%%%%%%%%%%%%%%%%%%%%%%%%%%%%%%%%%%%%%%%
\begin{eqnarray}
a\ &=& {\phantom{-}}1.29\ +\ 0.00250\ E\ -\ 1.76\ \times 10^{-6}\ E^2\ , 
\nonumber\\
\epsilon\ &=& -1.47\ -\ 0.00234\ E\ +\ 4.16\ \times 10^{-6}\ E^2\ ,
\label{Eps}
\end{eqnarray}
%%%%%%%%%%%%%%%%%%%%%%%%%%%%%%%%%%%%%%%%%%%%%%%%%%%%%%%%%%%%%%%%%%%%%%%%%%%
where the target energy E is in MeV. There was no conclusive evidence for a 
mass variation of them. With $a$ and $\epsilon$ so fixed, we then adjusted the
values of $l_0$ in each case so that actual measured neutron total 
cross-section data were fit using Eq.~(\ref{Fnform}).  Numerical values for
$l_0$ from that process are presented in a table in Ref.~\cite{De04}.
%%%%%%%%%%%%%%%%%%%%%%%%%%%%%%%%%%%%%%%%%%%%%%%%%%%%%%%%%%%%%%%%%%%%%%%%%%%%
The values of  $l_0$ increase monotonically with both mass and energy and that
is most evident in Fig.~\ref{l0vsE}, where the optimal values $l_0(E)$ are 
presented as different patterned and colored lines. The set for each of the  
masses 
(from 6 to 238) are given by those that increase in value respectively at 600
MeV. While that is obvious for most cases, note that there is some degree of
overlap in the values for ${}^{197}$Au  and for ${}^{208}$Pb.  
%%%%%%%%%%%%%%%%%%%%%%%%%%%%%%%%%%%%%%%%%%%%%%%%%%%%%%%%%%%%%%%%%%%%%%%%%%%%
%%%%%%%%%%%%%%%%%%%%%%%%%%%%%%%%%%%%%%%%%%%%%%%%%%%%%%%%%%%%%%%%%%%%%%%%%%%%%

%%%%%%%%%%%%%%%%%%%%%%%%%%%%%%%%%%%%%%%%%%%%%%%%%%%%%%%%%%%%%%%%%%%%%%%%%%%%

\begin{figure}
\centering
\scalebox{0.7}{\includegraphics{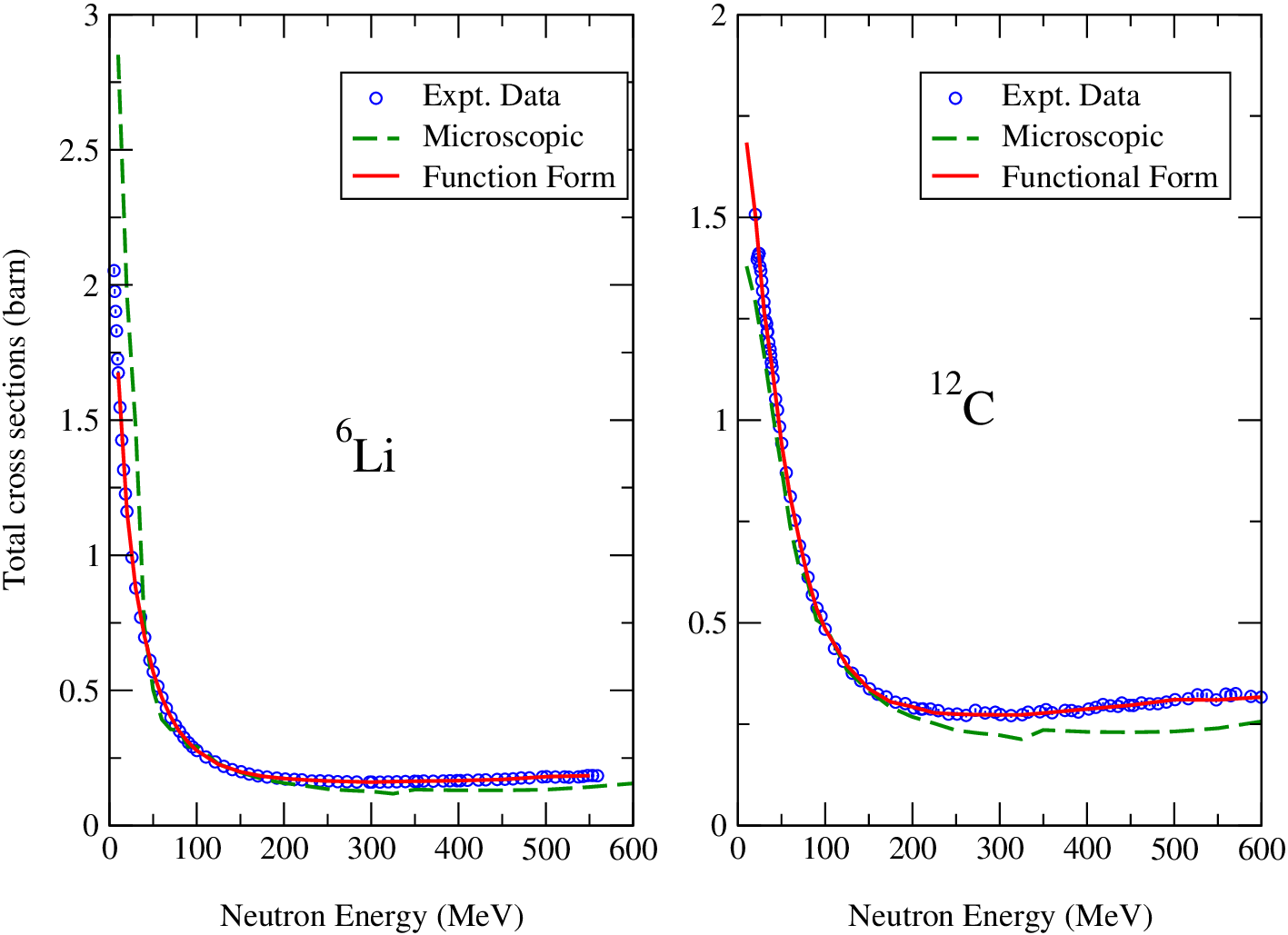}}
\caption{\label{li6c12-nX}
Total cross sections for neutrons scattered from $^6$Li (left) and  $^{12}$C
(right).}
\end{figure}

\begin{figure}
\centering
\scalebox{0.7}{\includegraphics{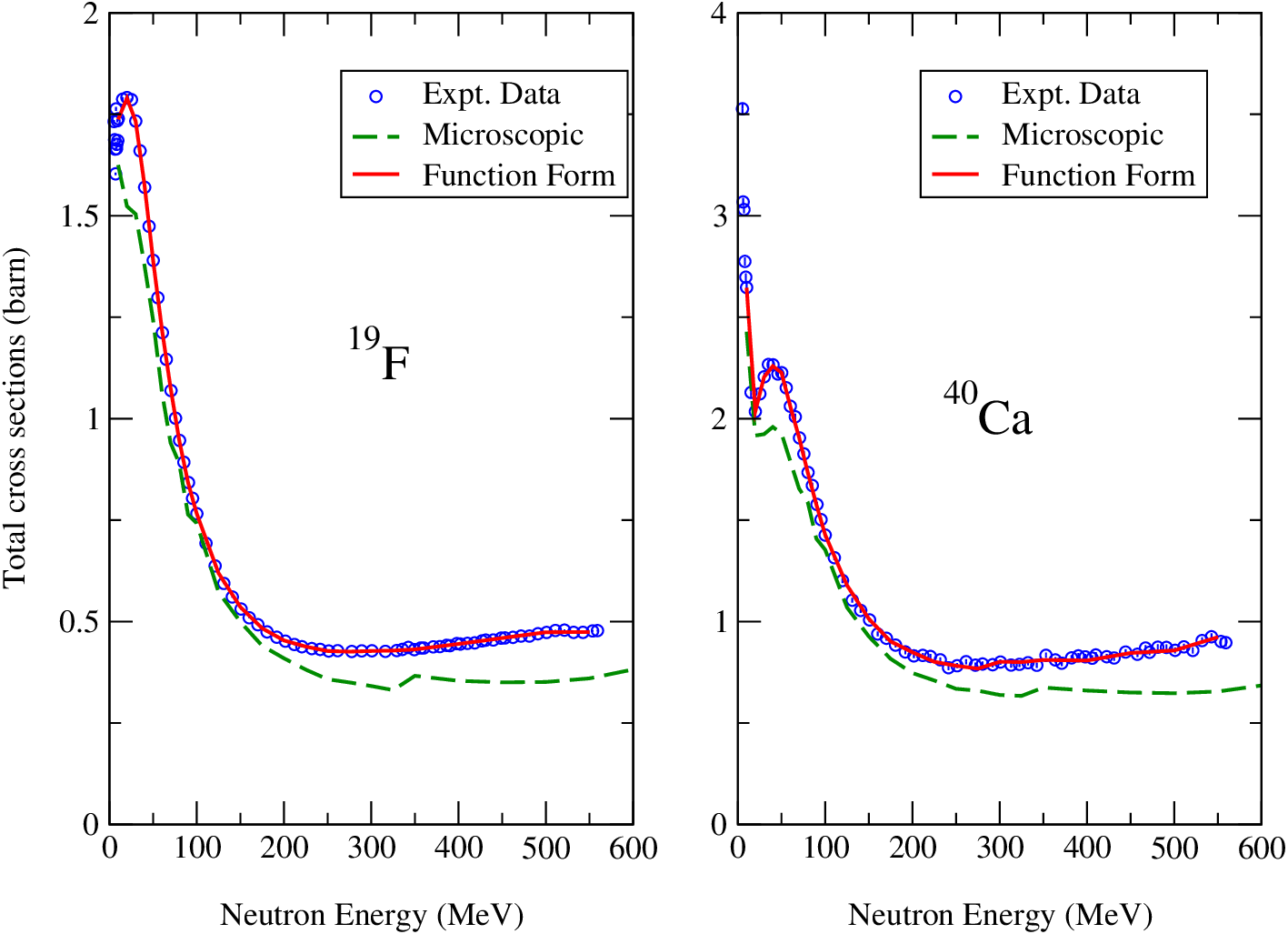}}
\caption{\label{f19ca40-nX}
Total cross sections for neutrons scattered from $^{19}$F (left) and  $^{40}$Ca
(right).}
\end{figure}

\begin{figure}                                                                  \centering
\scalebox{0.7}{\includegraphics{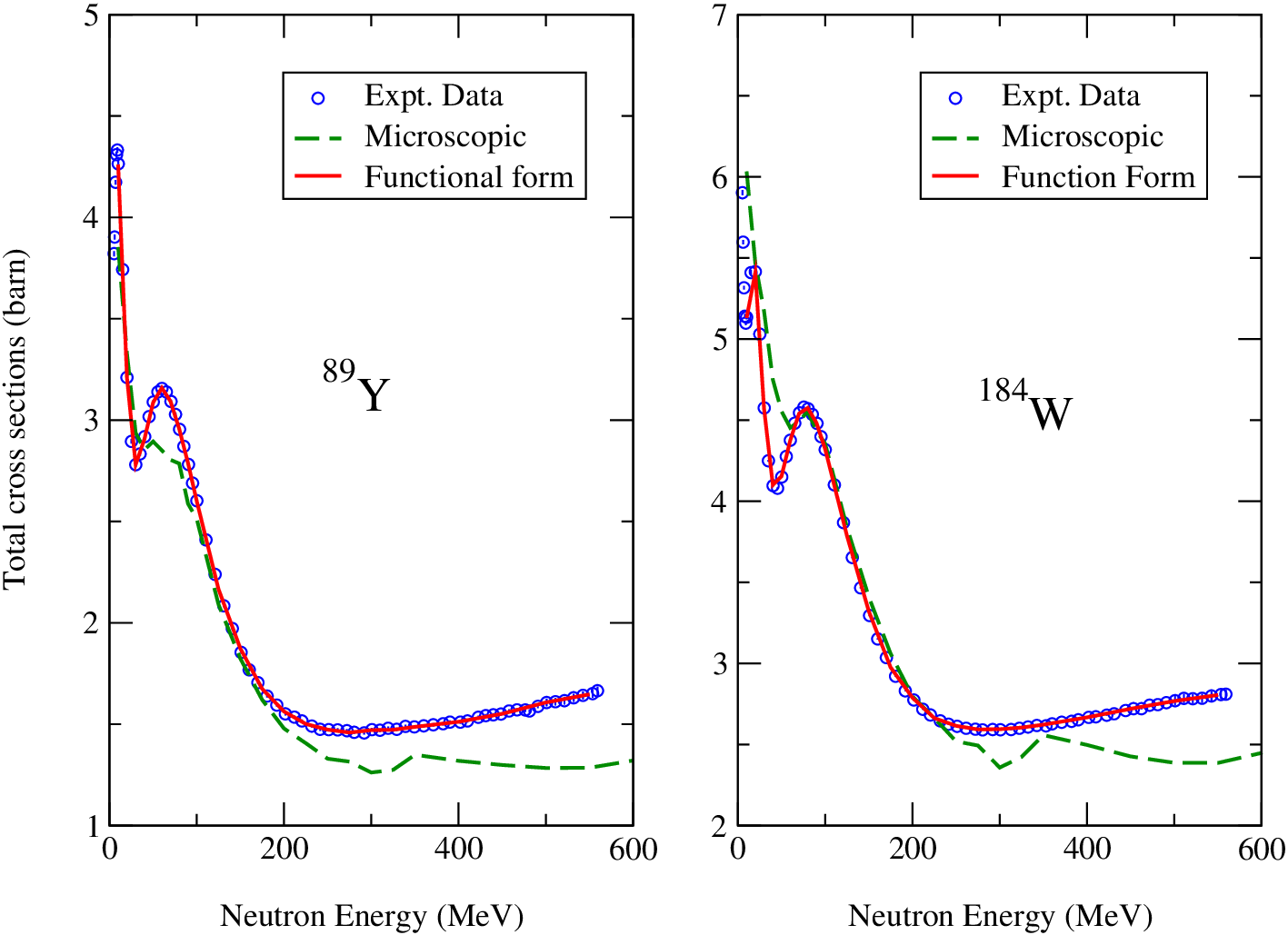}}
\caption{\label{y89w184-nX}
Total cross sections for neutrons scattered from $^{89}$Y (left) and $^{184}$W
(right).}
\end{figure}

\begin{figure}                                                                  \centering
\scalebox{0.7}{\includegraphics{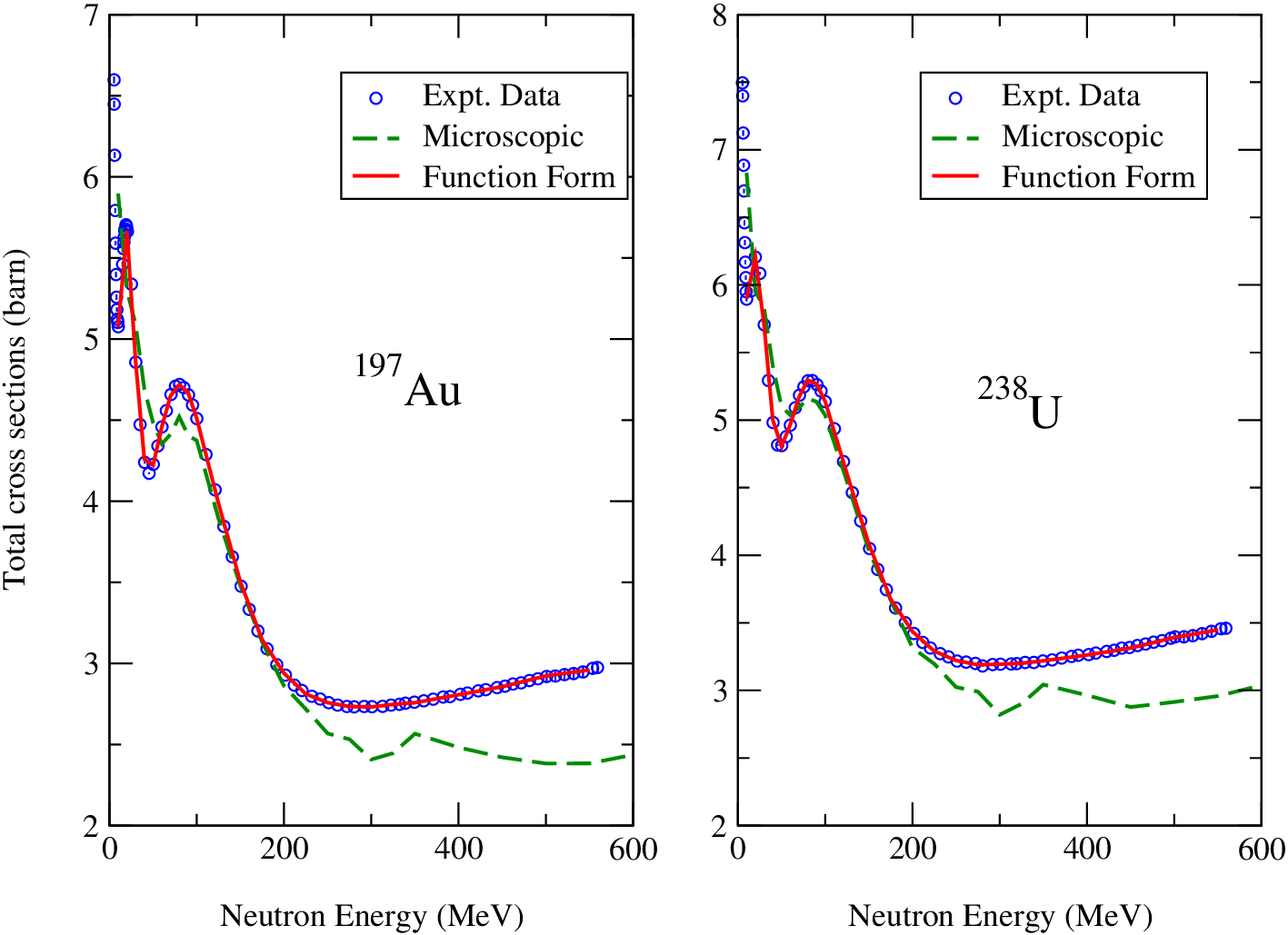}}
\caption{\label{au197u238-nX}
Total cross sections for neutrons scattered from $^{197}$Au (left) and 
$^{238}$U (right).}
\end{figure}

\begin{figure}                                                                  \centering
\scalebox{0.7}{\includegraphics{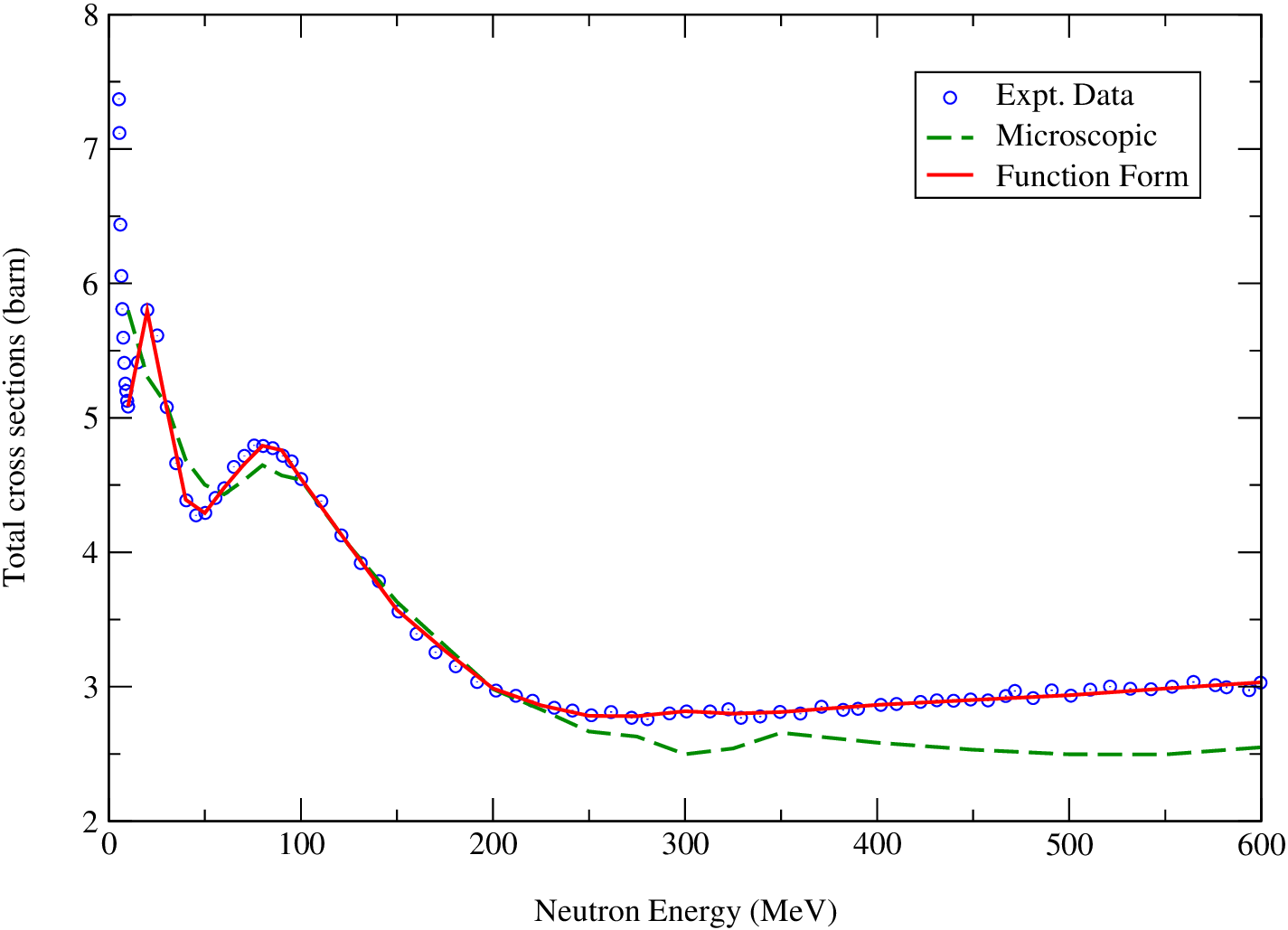}}
\caption{\label{pb208-nX}
Total cross sections for neutrons scattered from $^{208}$Pb.}
\end{figure}

The total neutron scattering cross sections generated using the function form
for partial total cross sections with the tabled values of $l_0$
and the energy function forms of Eq.~(\ref{Eps}) for $a$ and $\epsilon$,
are shown in Figs.~\ref{li6c12-nX}, \ref{f19ca40-nX}, \ref{y89w184-nX}, 
\ref{au197u238-nX}, and \ref{pb208-nX}.
They are displayed by the solid red lines that closely match the data which 
are portrayed by blue open  circles. The data that was taken from a survey by 
Abfalterer~{\it et al.}~\cite{Abf01} which includes data measured at LANSCE 
that are supplementary and additional to those published earlier by 
Finlay~{\it et al.}~\cite{Fin93}. For comparison we show results obtained from 
calculations made using $g$-folding optical potentials~\cite{Amos02}. Dashed 
green  lines represent the predictions obtained from those microscopic optical 
potential calculations. Clearly for energies 300 MeV and higher, those 
predictions fail.
%%%%%%%%%%%%%%%%%%%%%%%%%%%%%%%%%%%%%%%%%%%%%%%%%%

Predictions of the total cross sections  for neutrons scattered  from
$^6$Li and  $^{12}$C are compared to the data in Fig.~\ref{li6c12-nX} and those from  $^{19}$F and $^{40}$Ca are compared to the data in
 Fig.~\ref{f19ca40-nX}.
%%%%%%%%%%%%%%%%%%%%%%%%%%%%%%%%%%%%%%%%%%%%%%%%%%%%%%%%%%%%%%%%%
%%%%%%%%%%%%%%%%%%%%%%%%%%%%%%%%%%%%%%%%%%%%%%%%%%%%%%%%%%%%%%%%%%%%
%%%%%%%%%%%%%%%%%%%%%%%%%%%%%%%%%%%%%%%%%%%%%%%%%%%%%%%%%%%%%%%%%%%
Dashed green lines represent the results obtained from full folding 
optical potential
calculations and the solid red lines show the results obtained from functional
form calculations. In Fig.~\ref{li6c12-nX},  $g$ folding results for $^6$Li 
reflect the observed data
well from 30 MeV to 600 MeV, but overpredicted for the energies lower than
30 MeV. For $^{12}$C, $g$ folding results reproduce the data very well
for the energy range from 30 MeV to 200 MeV, but underpredicted for the
energies lower than 30 MeV and higher than 200 MeV. For $^{19}$F and
$^{40}$Ca cases, in Fig.~\ref{f19ca40-nX}, data fits well with the $g$-folding 
results for the energy
range 100 MeV 200 MeV, but under predicted for less than 100 MeV and
greater than 200 MeV. The peaks at 30 MeV for $^{19}$F and at 40 MeV for
$^{40}$Ca are underpredicted by the $g$-folding results. Excellent
reproductions of data made by functional form results for all cases.
%%%%%%%%%%%%%%%%%%%%%%%%%%%%%%%%%%%%%%%%%%%%%%%%%%%%%%%%%%%%%%%%%%%%

Our predictions for  the total cross sections for neutrons scattering
from $^{89}$Y, and  $^{184}$W are compared to the data in Fig.~\ref{y89w184-nX}
and those for $^{197}$Au, and $^{238}$U are compared to the data
in Fig.~\ref{au197u238-nX}.
%%%%%%%%%%%%%%%%%%%%%%%%%%%%%%%%%%%%%%%%%%%%%%%%%%%%%%%%%%%%%%%%%%%%
%%%%%%%%%%%%%%%%%%%%%%%%%%%%%%%%%%%%%%%%%%%%%%%%%%%%%%%%%%%%%%%%%%%%%%
%%%%%%%%%%%%%%%%%%%%%%%%%%%%%%%%%%%%%%%%%%%%%%%%%%%%%%%%%%%%%%%%%%%%%%%
In Fig.~\ref{y89w184-nX}, for $^{89}$Y, results predicted from $g$-folding 
calculations agree with the
experimental data
for the energy range between 10 MeV to 30 MeV and  90 MeV to 200 MeV, otherwise underpredicted.
For $^{184}$W, $g$-folding results agree with the data for the energy range
between 70 MeV to 250 MeV, but overpredict the data for the energy range
between 10 MeV to 70 MeV and underpredict for the energy above 250 MeV.
In Fig.~\ref{au197u238-nX}, for $^{197}$Au and $^{238}$U cases, 
$g$-folding results reflects the data
well for the energies below 200 MeV but overpredicted for the energies above
200 MeV. In both cases, the peaks near 40 MeV are overpredicted and peaks
near 80 MeV are underpredicted by the results obtained from $g$-folding
calculations. Functional form results are in excellent agreement with the
data in all cases.

%%%%%%%%%%%%%%%%%%%%%%%%%%%%%%%%%%%%%%%%%%%%%%%%%%%%%%%%%%%%%%%%%%%%%%%%%%%%%%
We show in Fig.~\ref{pb208-nX}, the results for neutron scattering 
from ${}^{208}$Pb. In this case we used Skyrme-Hartree-Fock model (SKM*) 
densities~\cite{Br00} to form the $g$-folding optical potentials. That 
structure when used to analyze proton and neutron scattering differential 
cross sections at 65 and 200 MeV gave quite excellent results~\cite{Ka02}.  
Indeed those analyzes were able to show selectivity for that SKM* model of 
structure and for the neutron skin thickness of 0.17 fm that it proposed.
%%%%%%%%%%%%%%%%%%%%%%%%%%%%%%%%%%%%%%%%%%%%%%%%%%%%%%%%%%%%%%%%%%%%%%%%%%%%
 
%%%%%%%%%%%%%%%%%%%%%%%%%%%%%%%%%%%%%%%%%%%%%%%%%%%%%%%%%%%%%%%%%%%%%%%%%%%%%
Using the SKM* model structure, the $g$-folding optical potentials gave the 
total cross sections shown by the dashed green curve in  
Fig.~\ref{pb208-nX}. Of 
all the results, we believe these for ${}^{208}$Pb point most strongly to a 
need to improve on the $g$-folding prescription as is used currently when 
energies are at and above pion threshold.  Nonetheless, it does do quite well
for lower energies, most notably giving a reasonable account of the 
Ramsauer resonances~\cite{Ko03} below 100 MeV. However, as with the other 
results, these $g$-folding values serve only to define a set of partial 
cross sections from which an initial guess at the parameter values of the 
function form is specified. With adjustment that form produces the solid 
 red curve
shown in Fig.~\ref{pb208-nX}, which is an excellent reproduction of 
the data,
as it was designed to do.  But the key feature is that the 
optimal fit parameter values still vary smoothly with mass and energy.

Without seeking further functional properties of the parameters, one could 
proceed as we have done this far but by using many more cases of target mass
and scattering energies so that a parameter tabulation as a data base may be 
formed with which any required value of total scattering cross section might 
be reasonably predicted (i.e. to within a few percent) by suitable 
interpolation on the data base, and the result used in Eq.~(\ref{Fnform}).
%%%%%%%%%%%%%%%%%%%%%%%%%%%%%%%%%%%%%
\section{Conclusions}

Total cross sections for 10 to 600 MeV neutron scattering from nuclei ranging
in mass from $^6$Li to $^{238}$U calculated by simple functional form are in 
excellent fit to the experimental data. We suggest the three parameter
function form for partial total cross sections that will give neutron total
cross sections without recourse to phenomenological optical potential 
parameter searches. That functional form also reproduces proton reaction cross
sections. The parameters that fit actual data show smooth trends with both 
energy and target mass. 
%%%%%%%%%%%%%%%%%%%%%%%%%%%%%%%%%%%%%%%%%%%%%%%%%%%%%%%%%%%%%%%%%%%%%%%%%

\begin{acknowledgments}
This research was supported by a research grant from the Australian Research 
Council and also by the National Science Foundation under Grant No. 0098645.
\end{acknowledgments}

\bibliography{ACNS04}

\end{document}